\documentclass[letterpaper, 10 pt, conference]{ieeeconf}  

\IEEEoverridecommandlockouts                              

\usepackage{booktabs}   
\usepackage{hyperref}
\usepackage{xcolor}     
\usepackage{graphicx}   
\usepackage{epsfig} 
\usepackage{mathptmx} 
\usepackage{times} 
\usepackage{amsmath} 
\usepackage{amssymb}  
\usepackage{siunitx}
\usepackage{tikz}
\usetikzlibrary{shapes, arrows.meta, positioning, calc, backgrounds, fit, shadows}

\usepackage{pgfplots}
\usepgfplotslibrary{fillbetween}
\pgfplotsset{compat=1.18}

\title{\LARGE \bf
End-to-End Voice Intent Recognition for Spontaneous Human-Drone Interaction with Naive Users}

\author{Allan Henry$^{1,2,3}$, Solange Rossato$^{1}$, Christian Graff$^{2}$, Sylvain Huet$^{3}$, Jose-Ernesto Gomez-Balderas$^{3}$%
\thanks{$^{1}$LIG, Univ. Grenoble Alpes, Grenoble, France}%
\thanks{$^{2}$LPNC, Univ. Grenoble Alpes, Grenoble, France}%
\thanks{$^{3}$GIPSA-lab, Univ. Grenoble Alpes, Grenoble, France}%
\thanks{{\tt\small firstname.lastname@univ-grenoble-alpes.fr}}%
}

\begin{document}

\maketitle
\thispagestyle{empty}
\pagestyle{empty}

\begin{abstract}
Voice control offers an intuitive alternative to manual drone piloting, yet most existing systems rely on rigid command vocabularies that fail to handle the spontaneous, disfluent speech of naive users. This paper addresses this gap by proposing an End-to-End Spoken Language Understanding architecture for real-time human-drone interaction in French. Our model combines a frozen Self-Supervised Learning acoustic encoder with a lightweight LSTM-based classification head, augmented by a cross-modal knowledge distillation objective that aligns acoustic representations with semantic embeddings from a text teacher, without requiring transcription at inference time. We evaluate our approach on VoiceStick, a novel French corpus of spontaneous speech collected during real teleoperation sessions with 29 nonexpert dyads. On simple voice commands, our best configuration achieves 93\% accuracy at 7\,ms inference latency, outperforming cascade baselines (79\%, 202\,ms) with a 29$\times$ speedup. On the full spontaneous speech test set, our architecture reaches 82\% accuracy, with cross-modal distillation consistently improving robustness across all configurations. These results demonstrate that End-to-End architectures are not only feasible but preferable for spontaneous voice-guided UAV teleoperation, combining semantic robustness, low latency, and calibrated confidence.
\end{abstract}

\section{Introduction}

The growing integration of drones into daily and industrial life requires the design of safer and more intuitive interaction methods. Within the field of Human-Robot Interaction (HRI), Human-Drone Interaction is emerging as a discipline in its own right \cite{tezzaStateArtHumanDrone2019, cauchardEmoDroneHRI16}. In this context, voice commands, which constitute an essential subdomain of Natural User Interfaces, have established themselves as a privileged mode of communication between humans and machines \cite{tezzaStateArtHumanDrone2019, wangMultimodalHumanRobot2024}.

The benefits of voice interaction for drone piloting are manifold. Unlike traditional manual controllers such as joystick-based remote controls, which require specific training and impose a significant cognitive load, the voice modality offers a markedly more accessible approach \cite{tezzaStateArtHumanDrone2019, wangMultimodalHumanRobot2024, thomasDesignDevelopmentAndroid2019, chengDesignNoviceFriendlyDrone2024, huangFlightCameraAction2019}. It enables intuitive control aligned with natural human communication, promotes hands-free operation, and substantially reduces the learning curve for operators \cite{alkasimEnhancingVoiceControlledDrone2026}. This accessibility is particularly critical for nonexpert users, who constitute the growing population of civilian drone operators \cite{FAAAerospaceForecast, ellenriederDesignAcceptanceIntuitive2023}.

However, the majority of voice control systems for drones have historically relied on strict, predefined command lists \cite{thomasDesignDevelopmentAndroid2019, choutriMultiLingualSpeechRecognitionBased2022, parkVehicleSpeechRecognition2023, poncelaCommandbasedVoiceTeleoperation2015, barbosaVoiceCommandedSystem2021, fayjieVoiceEnabledSmart2017, contrerasUnmannedAerialVehicle2020, yapiciogluVoiceCommandRecognition2021}. Users are constrained to employ a rudimentary and formatted vocabulary such as "take off", "land", or "move forward". While this keyword-based or rigid grammar-based approach \cite{rajapakshaResponsiveDroneAutopilot2019} enables basic functional control, it lacks flexibility and severely limits interaction fluidity.

To overcome these limitations and make drone guidance genuinely intuitive for novice users, recent research has moved toward the use of spontaneous speech and natural language \cite{atanovIntelligentVoiceControl2023}. Allowing an inexperienced user to formulate unstructured requests without having to memorize an imposed syntax makes interaction considerably more natural.

The development and evaluation of such systems are hindered, however, by the nature of available training data. The definition of what constitutes a "command" ranges from simple directional instructions (\cite{choutriMultiLingualSpeechRecognitionBased2022, fayjieVoiceEnabledSmart2017, contrerasUnmannedAerialVehicle2020, thomasDesignDevelopmentAndroid2019, yapiciogluVoiceCommandRecognition2021}) to highly precise languages such as military terminology \cite{parkVehicleSpeechRecognition2023, poncelaCommandbasedVoiceTeleoperation2015}, yet dominant public corpora are generally limited to isolated, out-of-context words \cite{wardenSpeechCommandsDataset2018, poirierEfficientSelfAttentionModel2023}. Furthermore, the literature relies heavily on English-language datasets. Although there is a growing need for multilingual interaction, research faces a well-documented shortage of French-language corpora specifically dedicated to spontaneous voice commands in robotics \cite{poirierEfficientSelfAttentionModel2023}.

At the architectural level, the translation of spontaneous voice signals into command execution is structured around two main paradigms. The traditional cascade approach chains an Automatic Speech Recognition (ASR) system to transcribe audio, followed by a semantic analysis module: either NLP-based or, more recently, LLM-based \cite{choiSpeechGuidedDroneControl2025, sunTrustNavGPTModelingUncertainty2024a, shahNavigationLargeLanguage2023a}. While this approach allows each module to be optimized independently, it suffers from accumulated latency and the propagation of transcription errors into the semantic module \cite{serdyukEndtoendSpokenLanguage2018, avilaMultimodalAudiotextualArchitecture2023}. In response, the End-to-End (E2E) approach is gaining traction by processing audio directly \cite{qianSpeechlanguagePretrainingEndtoend2021}. This architecture eliminates text as an intermediate representation, thereby reducing latency and preserving paralinguistic information such as intonation and hesitation, offering an interaction dynamic far better suited to spontaneous speech \cite{vitaleRichSpeechSignal2024, luSpokenDialogSum2025}.

The development of End-to-End SLU systems was historically hindered by the need for massive amounts of paired audio-to-intent data, which is computationally prohibitive and practically unavailable for specific robotic applications \cite{qianSpeechlanguagePretrainingEndtoend2021}. This bottleneck has recently been overcome by a major technological leap in Self-Supervised Learning (SSL) for speech representation. By pre-training large foundation models, such as wav2vec 2.0 \cite{baevskiWav2vec20Framework2020} or HuBERT \cite{hsu2021hubert}, on tens of thousands of hours of unlabelled audio, these models learn highly robust acoustic and phonetic representations. This paradigm shift, heavily popularized by benchmarks like SUPERB \cite{yangSUPERBSpeechProcessing2021}, allows for the extraction of rich features using a frozen SSL encoder, upon which only a lightweight downstream classification head, such as a simple linear layer or a recurrent neural network like a LSTM, needs to be trained. This "frozen encoder" approach drastically reduces the need for large task-specific datasets \cite{yangSUPERBSpeechProcessing2021, aroraEvaluatingSFM2024}. Furthermore, to address the high computational and memory footprint of these massive foundation models, knowledge distillation techniques, such as DistilHuBERT \cite{changDistilhubertSpeechRepresentation2022}, have been introduced. Distillation compresses the model architecture, significantly reducing inference time by up to 73\% while preserving representation quality. Together, these advances make the End-to-End approach practically viable for the strict real-time latency and embedded constraints of UAV teleoperation.

However, despite its potential, the End-to-End approach remains largely underexplored in the context of spontaneous speech in robotics, due to a lack of suitable corpora. The data required for training (unstructured utterances produced by naive users in real interaction conditions) are virtually absent from the literature, and all the more so in French. Faced with these bottlenecks, we propose several contributions. First, we introduce VoiceStick, a dedicated French spontaneous speech corpus for human-drone interaction. The corpus collection protocol and linguistic analysis are described in detail in a companion paper currently under review. The key characteristics relevant to this evaluation are reported in Section \ref{sec:dataset}. Second, we propose an End-to-End SLU architecture combining a frozen SSL encoder with a lightweight classification head, achieving competitive intent recognition accuracy with a latency compatible with real-time UAV control. Third, we provide a systematic comparison against cascade baselines, demonstrating that our approach better handles the disfluencies and linguistic variability inherent to spontaneous speech from nonexpert users.

\section{Proposed Method}
\label{sec:method}

We propose a streamlined End-to-End architecture designed to minimize latency while maintaining robustness to spontaneous speech. To compensate for the lack of textual supervision, we introduce a cross-modal training strategy based on knowledge distillation.

\subsection{Model Architecture}

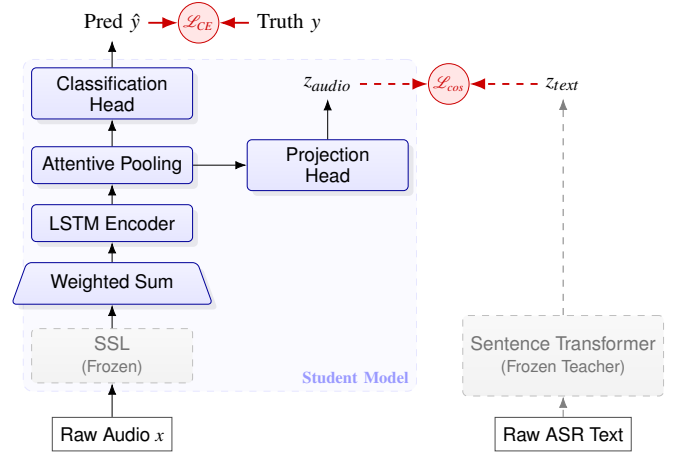
\begin{figure}[ht]
\centering
\resizebox{\columnwidth}{!}{%
\begin{tikzpicture}[
    font=\small,
    >=LaTeX,
    node distance=0.4cm and 0.6cm, 
    student/.style={
        draw=blue!60!black, top color=blue!5, bottom color=blue!10, 
        rounded corners=2pt, 
        minimum height=0.55cm, 
        minimum width=2.4cm, 
        align=center, drop shadow={opacity=0.15}, font=\sffamily\footnotesize
    },
    teacher/.style={
        draw=gray!50, top color=gray!5, bottom color=gray!10, 
        dashed, rounded corners=2pt, 
        minimum height=0.55cm, 
        minimum width=2.4cm, 
        align=center, text=gray!90!black, font=\sffamily\footnotesize
    },
    trapez/.style={
        student, trapezium, trapezium left angle=70, trapezium right angle=70, 
        minimum width=3.0cm 
    },
    input/.style={
        draw=black!70, fill=white, minimum height=0.5cm, minimum width=1.6cm,
        align=center, font=\sffamily\footnotesize
    },
    loss/.style={
        circle, draw=red!80!black, fill=red!10, inner sep=1pt, 
        font=\scriptsize\bfseries, text=red!80!black
    },
    tensor/.style={font=\tiny\itshape, text=gray!70, inner sep=1pt}
]

    
    \node[input] (audio) {Raw Audio $x$};
    \node[teacher, above=0.5cm of audio] (w2v_frozen) {SSL \\ \scriptsize (Frozen)};
    \node[trapez, above=0.35cm of w2v_frozen] (sum) {Weighted Sum};
    \node[student, above=0.3cm of sum] (lstm) {LSTM Encoder};
    \node[student, above=0.3cm of lstm] (pool) {Attentive Pooling};
    \node[student, above=0.4cm of pool] (cls_head) {Classification\\Head};
    \node[above=0.4cm of cls_head] (pred) {Pred $\hat{y}$};
    
    \node[right=1.5cm of pred] (gt) {Truth $y$};

    
    \node[student, right=0.8cm of pool] (dist_head) {Projection\\Head};
    \node[above=0.6cm of dist_head] (z_audio) {$z_{audio}$};

    
    \path (dist_head.east) ++(0.8cm, 0) coordinate (bert_x);
    \node[teacher, minimum height=1.2cm] (bert) at (bert_x |- w2v_frozen.center) [anchor=west] {Sentence Transformer\\ \scriptsize (Frozen Teacher)};
    \node[input] (text) at (bert |- audio) {Raw ASR Text};
    \node (z_text) at (bert |- z_audio) {$z_{text}$};


    \draw[->] (audio) -- (w2v_frozen);
    \draw[->] (w2v_frozen) -- (sum);
    \draw[->] (sum) -- (lstm);
    \draw[->] (lstm) -- (pool);
    \draw[->] (pool) -- (cls_head);
    \draw[->] (cls_head) -- (pred);
    
    \draw[->] (pool.east) -- node[above, tensor] {} (dist_head.west);
    \draw[->] (dist_head) -- (z_audio);

    \draw[->, dashed, gray] (text) -- (bert);
    \draw[->, dashed, gray] (bert) -- (z_text);


    \path (pred) -- node[loss] (l_ce) {$\mathcal{L}_{CE}$} (gt);
    \draw[->, thick, red!80!black] (pred) -- (l_ce);
    \draw[->, thick, red!80!black] (gt) -- (l_ce);

    \path (z_audio) -- node[loss] (l_cos) {$\mathcal{L}_{cos}$} (z_text);
    \draw[->, dashed, thick, red!80!black] (z_audio) -- (l_cos);
    \draw[->, dashed, thick, red!80!black] (z_text) -- (l_cos);

    \begin{scope}[on background layer]
        \node[fit=(w2v_frozen)(cls_head)(dist_head), 
              fill=blue!2, draw=blue!10, rounded corners, dashed,
              label={[blue!40, font=\bfseries\scriptsize, anchor=south east]south east:Student Model}] {};
    \end{scope}
    \useasboundingbox (current bounding box.south west) rectangle ([yshift=0.5cm]current bounding box.north east);
\end{tikzpicture}
}
\caption{Proposed End-to-End architecture.}
\label{fig:arch}
\vspace{-0.6cm} 
\end{figure}

Our architecture, depicted in Fig.~\ref{fig:arch} extracts acoustic features using a pre-trained Self-Supervised Learning (SSL) module based on the \textit{Wav2Vec} framework \cite{baevskiWav2vec20Framework2020}. Since the choice of pre-training data is a primary factor in model robustness, we investigate several French and multilingual checkpoints (see Section~\ref{sec:results}). Rather than relying solely on the final layer, we compute a learnable weighted sum of the hidden states from the last $N=4$ Transformer layers to capture a richer set of phonetic and semantic information. To ensure computational efficiency and mitigate catastrophic forgetting, we freeze the entire backbone.

To handle the significant variance in duration and pacing characteristic of spontaneous speech, we implement a multi-stage aggregation process. First, an LSTM models long-range sequential dependencies across the frame-level feature sequence. An \textit{Attentive Pooling} mechanism then collapses the LSTM outputs into a single segment-level representation, effectively weighting the most informative temporal positions while filtering out silence and irrelevant environmental noise. While Transformer-based architectures were also evaluated for this sequence modeling task, they yielded no significant accuracy gains while increasing computational overhead. Consequently, the LSTM was retained to prioritize minimal inference latency without compromising performance. The resulting pooled representation, denoted as $h_{audio}$, serves as the central shared representation for the dual-head objectives described in the following section.

\subsection{Cross-Modal Knowledge Distillation}
\label{ssec:distillation}

To bridge the gap between acoustic signals and semantic intent, we employ a multi-task learning strategy within a Teacher-Student framework. The architecture branches into two parallel heads: a Classification Head that applies a linear layer to $h_{audio}$ to predict command probabilities $\hat{y}$, and a Distillation Head that projects $h_{audio}$ into a semantic vector $z_{audio}$ designed to match the latent space of the text encoder. We adopt a single-label classification framework, motivated by the sequential nature of spontaneous drone guidance commands: prior analysis of such interactions demonstrated that 98.6\% of utterances follow a single-direction structure \cite{henryVoiceCommandsGuidance2025}, justifying this design choice without significant loss of operational coverage.

During training, a frozen \textit{Sentence Transformer} \cite{reimersSentenceBERTSentenceEmbeddings2019} (the Teacher) encodes the transcript into a target semantic embedding $z_{text}$. We optimize a joint loss combining cross-entropy for intent classification and cosine distance for semantic alignment: $\mathcal{L}_{total} = \mathcal{L}_{CE}(\hat{y}, y) + \mathcal{L}_{cos}(z_{audio}, z_{text})$, where $\mathcal{L}_{CE}$ optimizes the recognition accuracy using ground-truth labels $y$, and $\mathcal{L}_{cos}$ minimizes the cosine distance between the student's projection $z_{audio}$ and the teacher's embedding $z_{text}$. This auxiliary objective forces the acoustic encoder to capture linguistic structures implicit in the text. At inference time, only the student branch is active, with the classification head producing the final intent prediction. This architectural design is motivated by the operational constraints of real-time HRI: the frozen backbone minimizes inference latency, while the distillation objective ensures semantic robustness without requiring textual transcription at deployment time.

\subsection{Encoders and Training}
To validate our approach, we benchmark several text teacher configurations, ranging 
from French-specialized models (CamemBERT-Large \cite{martinCamemBERTTastyFrench2020}) 
to multilingual encoders \cite{wangTextEmbeddingsWeaklySupervised2024, 
chenM3EmbeddingMultiLingualityMultiFunctionality2025} and lightweight distilled 
variants \cite{wangMiniLMDeepSelfAttention2020}. On the acoustic side, we compare 
multilingual Wav2Vec backbones \cite{conneauUnsupervisedCrosslingualRepresentation2020, 
hsu2021hubert, chen2022wavlm, baevski2022data2vec, wangVoxPopuliLargeScaleMultilingual2021} 
against French-adapted versions \cite{grosman2021xlsr53-large-french, 
parcolletLeBenchmark20Standardized2024, lePantagruelUnifiedSelfSupervised2026} to assess 
the impact of language specialization. The full comparison is reported in Table~I. 
All models are fine-tuned using AdamW (lr $= 10^{-3}$, weight decay $= 10^{-4}$, 
batch size $= 1$) with early stopping (patience $= 5$) on a single \textit{RTX 2000 Ada} GPU. 
This end-to-end approach is compared against a \textit{Cascade System} baseline, 
which chains \textit{Whisper Large-v2} \cite{radfordRobustSpeechRecognition} with a 
Sentence-Transformer for intent classification.

\subsection{Dataset: VoiceStick Corpus}
\label{sec:dataset}
To evaluate our approach in realistic HRI conditions, we use the VoiceStick corpus, a novel French dataset collected to investigate spontaneous speech in human-drone teleoperation. It consists of approximately 2 hours of audio recordings, comprising 4219 utterances across 2421 unique sentences and a vocabulary of 669 unique words. Unlike read-speech command datasets, this corpus captures the hesitations, disfluencies, and prosodic urgency inherent to real-time teleoperation. The corpus will be made publicly available upon acceptance of this work.

The corpus was collected from 29 native French-speaking student dyads (46 female and 12 male, mean age 20.1, std 1.99 years), none of whom had prior experience with drone piloting. The task was asymmetric by design: the Pilot (Room A) controlled a commercial micro-drone via its FPV camera feed to reach invisible 3D targets, while the Guide (Room B) monitored a virtual twin of the environment reconstructed via a motion capture system, and provided vocal instructions to direct them. Utterances range from terse explicit commands such as \textit{"Avance"} (\textit{Forward}) or \textit{"Monte un peu"} (\textit{Go up a bit}), to disfluent, context-dependent instructions such as \textit{"Euh... vas à gauche, non à droite"} (\textit{Uh... go left, no right}) or \textit{"Tourne, tourne, voilà stop"} (\textit{Turn, turn, okay stop}). A full description of the corpus, including collection protocol and linguistic analysis, is available in \cite{henryVoiceStickTALN2026} and the dataset is publicly accessible at \url{https://zenodo.org/records/19882638}. The evaluation is conducted on manually annotated dyads comprising 412 utterances, ensuring no overlap with the training distribution.

Unlike standard ASR corpora, each utterance carries dual labels: the \textit{Pilot Action} (derived from time-aligned joystick inputs) as pragmatic ground truth, and the \textit{Semantic Intent} from manual transcription. The Pilot Actions show strong class imbalance with ``Forward'' predominating (approx.\ 33\%), across 7 unified classes: \textit{Forward, Backward, Up, Down, Left, Right,} and \textit{No Command}.

Our evaluation protocol relies on two distinct test splits to disentangle acoustic performance from semantic ambiguity. First, an \textit{Explicit} subset ($N=250$) focuses on unambiguous commands (e.g., \textit{"Avance" (Forward)}, \textit{"Monte un peu" (Go up a bit)}), where the Ground Truth is derived strictly from the textual transcript to isolate the model's acoustic understanding. Second, the \textit{Full} test set ($N=412$) encompasses both explicit and spontaneous utterances; in this case, the Ground Truth is defined by the pilot's actual control input. 

Prior to model evaluation, the test set underwent a manual re-annotation pass to address three categories of inherent ambiguity in the pilot ground truth. First, pilot execution errors (16.1\% of explicit commands) were identified as cases where the guide issued a clear directional instruction but the pilot executed a different movement. Second, lateral inversions (13.5\% of spontaneous utterances) correspond to turn commands where either Left or Right would be valid given the 3D environment, including cases where pilots confused their own lateral orientation. Third, parallel commands (2.9\%) involve utterances containing two explicit directions, where both the pilot and a model predicting the other half of the instruction can be considered correct. This corrected annotation serves as the reference ground truth for all Full test set evaluations.

Post-experiment questionnaires confirm that dyads rated the interaction as comfortable ($W = 0.0$, $p < 0.001$) and effortless ($W = 3.5$, $p < 0.001$).

\section{Results}
\label{sec:results}

\subsection{Performance on the \textit{Explicit} Subset}
Table~\ref{tab:main_results} summarizes the comparative results between our proposed End-to-End architecture, Text-Based Oracles, and standard Cascade baselines. Our analysis focuses on two metrics critical for real-time control: semantic accuracy on the \textit{Explicit} subset and inference latency.

The \textit{Text-Based Oracle} establishes a semantic upper bound of 96~\% accuracy using manual transcripts at a latency of \SI{22}{\milli\second}. The practical \textit{Cascade} baseline (\textit{Whisper} + \textit{CamemBERT-Large}) achieves 79~\% accuracy with a latency of \SI{202}{\milli\second}. A McNemar test confirms that the performance gap between the best E2E configuration and the cascade is statistically significant ($p < 0.001$), indicating that the E2E architecture systematically recovers utterances where the cascade fails, with no instance where both models fail simultaneously.

Among the proposed End-to-End models, the French-specialized \textit{XLSR-53-FR} achieves 93~\% accuracy at \SI{7}{\milli\second} (\SI{106}{\milli\second} on CPU), \textit{WavLM} reaches 92~\% at \SI{14}{\milli\second}, while the generic multilingual \textit{XLSR-53} obtains only 68~\%. The remaining backbones range from 83~\% to 94~\% at 10--14\,ms, with an average of 87~\% at \SI{11}{\milli\second}.

The ablation study (Table \ref{tab:main_results}, bottom) reports performance when disabling the semantic teacher (distillation loss removed). \textit{XLSR-53-FR} falls from 93\% to 82\% and \textit{WavLM} from 92\% to 87\%. The remaining configurations show similar degradation patterns. The average accuracy without distillation is 80\%, compared to 87\% with distillation.

This constitutes a direct ablation of the distillation objective: removing it degrades performance by an average of 7 percentage points across all configurations, with the largest drop observed for \textit{XLSR-53-FR} ($-$11~pp) and \textit{WavLM} ($-$5~pp).

\begin{table}[ht]
    \centering
    \caption{Performance comparison on the Explicit subset. \textit{Oracle} uses manual transcripts. \textit{Cascade} uses ASR output. \textit{End-to-End} uses raw audio. For each category, we report representative models and the category average.}
    \label{tab:main_results}
    \resizebox{\linewidth}{!}{
    \begin{tabular}{lccc} 
        \toprule
        \textbf{Method} & \textbf{Acc.} &  \textbf{Acc.}& \textbf{Latency} \\
         & \textbf{Explicit} (\%) & \textbf{Full} (\%) & (ms) \\
        \midrule
        \multicolumn{4}{l}{\textit{\textbf{Text-Based Oracle (Input: Manual Text)}}} \\
        CamemBERT (Large) & \textbf{96} & \textbf{92} &\phantom{0}22  \\ 
        \cmidrule(lr){1-4}
        \textit{Average$^{\star}$} & \textit{92} & 86 & \textit{\phantom{0}16}  \\        
        \midrule
        \multicolumn{4}{l}{\textit{\textbf{Cascade Systems (Input: Audio $\rightarrow$ Whisper $\rightarrow$ Text)}}} \\
        CamemBERT (Large) & \textbf{79} & \textbf{59} & 202  \\
        \cmidrule(lr){1-4}
        \textit{Average$^{\star}$} & \textit{76} & 52 & \textit{196} \\
        \midrule
        \multicolumn{4}{l}{\textit{\textbf{Proposed End-to-End (Input: Raw Audio)}}} \\
        XLSR-53-FR (Large) & \textbf{93} & \textbf{82} &\phantom{00}\textbf{7}  \\
        XLSR-53 (Large)$^\dagger$ & 68 & 50 & \phantom{0}10  \\
        WavLM (Large) & 92 & 64 & \phantom{0}14 \\
        \cmidrule(lr){1-4}
        \textit{Average$^{\star}$} & \textit{87} & 64 & \textit{\phantom{0}11}  \\
        \midrule
        \multicolumn{4}{l}{\textit{\textbf{End-to-End without Distillation (Input: Raw Audio)}}} \\
        XLSR-53-FR (Large) & 82 & \textbf{65} & \phantom{00}\textbf{7}  \\
        WavLM (Large) & \textbf{87} & 59 & \phantom{0}14  \\
        \cmidrule(lr){1-4}
        \textit{Average$^{\star}$} & \textit{80} & 52 & \textit{\phantom{0}11} \\
        \bottomrule
        \multicolumn{4}{l}{\footnotesize $^\dagger$Excluded from ablation due to low baseline performance.} \\
        \multicolumn{4}{l}{\footnotesize $^{\star}$Average computed over all tested configurations.} \\
    \end{tabular}
    }
\end{table}

\subsection{Performance on the Full Set}
When extending to the manually re-annotated Full test set, our E2E architecture achieves 82\% accuracy. For reference, evaluation against the raw pilot ground truth yields 63\% across all architectures, including the Text-Based Oracle, confirming that the performance gap stems from ground truth noise rather than model failures.

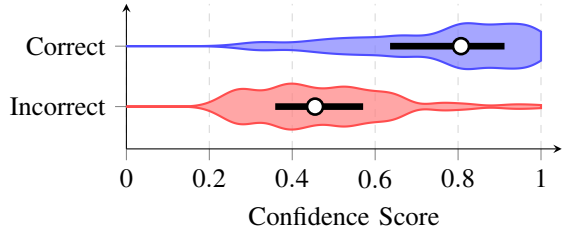
\begin{figure}[ht]
\centering
\begin{tikzpicture}
\begin{axis}[
    width=0.85\columnwidth,
    height=3.5cm,
    ymin=0.3, ymax=2.7,
    xmin=0, xmax=1.05,
    ytick={1, 2},
    yticklabels={Incorrect, Correct},
    xlabel={Confidence Score},
    xmajorgrids=true,
    grid style={dashed, gray!30},
    axis lines=left,
    tick align=outside,
    xtick={0, 0.2, 0.4, 0.6, 0.8, 1.0},
]

\addplot [name path=ct, draw=none, forget plot]
    table [y expr=2+\thisrow{density}*0.38, x=x] {kde_correct.dat};
\addplot [name path=cb, draw=none, forget plot]
    table [y expr=2-\thisrow{density}*0.38, x=x] {kde_correct.dat};
\addplot [fill=blue!50, fill opacity=0.7, draw=blue!70, thick]
    fill between [of=ct and cb];

\addplot [black, line width=2.5pt, forget plot]
    coordinates {(0.636, 2) (0.912, 2)};
\addplot [only marks, mark=*, mark size=3pt,
    mark options={fill=white, draw=black, line width=1pt}, forget plot]
    coordinates {(0.807, 2)};

\addplot [name path=it, draw=none, forget plot]
    table [y expr=1+\thisrow{density}*0.38, x=x] {kde_incorrect.dat};
\addplot [name path=ib, draw=none, forget plot]
    table [y expr=1-\thisrow{density}*0.38, x=x] {kde_incorrect.dat};
\addplot [fill=red!50, fill opacity=0.7, draw=red!70, thick]
    fill between [of=it and ib];

\addplot [black, line width=2.5pt, forget plot]
    coordinates {(0.359, 1) (0.571, 1)};
\addplot [only marks, mark=*, mark size=3pt,
    mark options={fill=white, draw=black, line width=1pt}, forget plot]
    coordinates {(0.455, 1)};

\end{axis}
\useasboundingbox (current bounding box.south west) rectangle ([yshift=0.5cm]current bounding box.north east);
\end{tikzpicture}
\caption{Confidence score distributions for correct and incorrect predictions
(Mann-Whitney $U=26333$, $p=4.35 \times 10^{-27}$).
Bars indicate IQR, white dots indicate median.}
\label{fig:confidence}
\vspace{-0.6cm} 
\end{figure}

The model demonstrates a statistically significant distinction in confidence 
distributions between correct and incorrect predictions (Fig.~\ref{fig:confidence}): 
the median confidence for correct predictions is 0.81, compared to 0.46 for 
misclassified instances (Mann-Whitney $U=26333$, $p=4.35 \times 10^{-27}$).

\section{Discussion}
The results validate the core motivation of this work: while traditional cascade systems focus on transcription accuracy, drone piloting imposes strict real-time constraints where delays can directly compromise flight stability and mission success. UAV teleoperation systems require end-to-end latencies below \SI{100}{\milli\second} to maintain stable control \cite{brodnevsRequirementsEndtoEndDelays2021}. By mapping audio directly to intent, our End-to-End architecture reduces inference to \SI{7}{\milli\second}, well within this operational threshold, compared to \SI{202}{\milli\second} for the cascade baseline, a 29 $\times$ speedup. This is particularly critical when compared to current trends integrating Large Language Models \cite{choiSpeechGuidedDroneControl2025, sunTrustNavGPTModelingUncertainty2024a, shahNavigationLargeLanguage2023a}: while LLMs offer high semantic flexibility, their inference times of 4 to 8 seconds are fundamentally incompatible with UAV flight dynamics.

Critically, the performance gap between the cascade and E2E architectures cannot be attributed to ASR quality alone. Whisper Large-v2 achieves a WER of 6.64~\% on our corpus, suggesting that the 14-point accuracy gap (79~\% vs. 93~\%) stems from the cascade architecture itself rather than transcription errors. This is consistent with the known limitation of cascade systems, where error propagation from the ASR stage directly contaminates the semantic module \cite{serdyukEndtoendSpokenLanguage2018, avilaMultimodalAudiotextualArchitecture2023}. In contrast, the E2E approach operates directly on the acoustic signal, retaining sub-phonemic and prosodic information that text transcription discards by design \cite{vitaleRichSpeechSignal2024, luSpokenDialogSum2025}. Whether this specifically accounts for the observed accuracy gap, or whether it reflects broader regularities captured during SSL pre-training, remains an open question that future probing experiments could address.

The choice of acoustic backbone proves decisive for this robustness. The generic multilingual XLSR-53 backbone degrades significantly to 68~\% accuracy, failing to map French-specific phonological and prosodic patterns to semantic intents without the benefit of intermediate transcription. Conversely, the French-specialized XLSR-53-FR reaches 93~\%, demonstrating that language-specific pre-training is essential for direct audio-to-intent mapping on unscripted speech. This finding is consistent with the broader observation in the speech community that SSL models trained on language-specific data better capture the phonotactic and prosodic properties of spontaneous speech \cite{parcolletLeBenchmark20Standardized2024}.

The remaining backbones \cite{hsu2021hubert, baevski2022data2vec, parcolletLeBenchmark20Standardized2024, lePantagruelUnifiedSelfSupervised2026} yield an average accuracy of 87~\% with distillation and 80~\% without, confirming that distillation consistently improves performance across architectures regardless of pre-training origin.

When extending to the Full test set, the performance drop observed across all architectures, including the Text-Based Oracle (from 92\% to 80\%), illustrates the inherent difficulty of intent recognition in spontaneous speech, even with perfect transcription. This confirms a fundamental challenge inherent to real-world HRI evaluation: the ground truth itself is noisy. Human operators adapt, correct, and sometimes contradict their own spoken instructions in real time \cite{rauxDynamicsActionCorrections2010}, making raw pilot actions an imperfect reference signal. Under the manually corrected annotation, our architecture achieves 82\%, suggesting that a system consistently following spoken intent may be more operationally reliable than raw accuracy metrics indicate. 

A qualitative analysis of misclassified utterances reveals two dominant error patterns. First, confusion between \textit{Left} and \textit{Right} accounts for a disproportionate share of errors, consistent with the lateral inversion ambiguity identified during re-annotation: in one representative case, the utterance \textit{"à gauche"} (\textit{to the left}) was misclassified as \textit{Up} with a moderate confidence of 0.48, while \textit{Left} remained the second-ranked class at 0.27: acoustic analysis suggests the speaker produced the vowel in \textit{gauche} with a rounded quality phonetically close to the vowel in \textit{haut} (\textit{up}), a speaker-level variation that a text-based system would not encounter. Second, the boundary between \textit{Forward} and \textit{No Command} proves difficult to resolve acoustically, particularly for short, hesitant utterances such as \textit{"tout droit"} (\textit{straight ahead}), 
predicted as \textit{No Command} with confidence 0.53 
despite an unambiguous directional intent, likely due to 
insufficient acoustic context in a 0.39s segment. Crucially, the confidence score distribution (Fig.~\ref{fig:confidence}) provides a practical signal for these failure cases: 72.7\% of misclassified instances fall below a confidence threshold of 0.636 corresponding to the first quartile of the correct prediction 
confidence distribution, suggesting that such a rejection mechanism could substantially 
reduce operational errors without degrading throughput. This calls for rethinking evaluation protocols in HRI beyond simple input-output accuracy measures.

A current limitation concerns movement magnitude: the model classifies direction but not extent, e.g. \textit{"Vas un peu à gauche"} (\textit{Go slightly to the left}) vs. \textit{"Fais un grand écart à gauche"} (\textit{Move far to the left}). Prosodic and durational cues present in the acoustic signal could address this via an additional regression head.

\section{Conclusion}
This work demonstrates that an End-to-End SLU architecture combining a frozen SSL encoder with cross-modal knowledge distillation is both feasible and preferable for spontaneous voice-guided UAV teleoperation. Operating at 7\,ms inference latency with 93\% accuracy on unambiguous commands, it meets the real-time constraints of assistive HRI scenarios while remaining robust to the disfluencies that rigid grammars cannot handle. Although evaluated on French, the architecture is language-agnostic and extends naturally to any collaborative teleoperation scenario where a remote operator must guide a field agent in real time.

Future work will explore streaming inference for incremental prediction and 
prosody-aware movement amplitude estimation, building on the acoustic signal's 
durational and prosodic cues to jointly predict both intent and movement magnitude.

\section*{Acknowledgment}
This work is supported by the French National Research Agency in the framework of the "Investissements d'avenir" program (ANR-15-IDEX-02) and has been partially supported by ROBOTEX 2.0 (Grants ROBOTEX ANR-10-EQPX-44-01 and TIRREX ANR-21-ESRE-0015) funded by the French program Investissements d'avenir. This work has benefited from collaboration with members of SAMGuide (ANR-21-VE33-0011-01). The authors would like to thank all participants who took part in the data collection experiment.

\bibliographystyle{IEEEtran}
\bibliography{bibi}

\end{document}